\documentstyle[aps,prb,floats,twocolumn,epsf]{revtex} 

\begin{document}

\twocolumn[\hsize\textwidth\columnwidth\hsize\csname
@twocolumnfalse\endcsname

\title{Low-temperature properties
of Pb(Zr$_{1-x}$Ti$_{x}$)O$_{3}$ solid solutions \\
 near the morphotropic phase boundary}
 
\author{L. Bellaiche$^{1}$, 
        Alberto Garc\'{\i}a$^{2}$ and David Vanderbilt$^{3}$}

\address{$^{1}$ Physics Department,\\
                University of Arkansas, Fayetteville, Arkansas 72701, USA\\
         $^{2}$Departamento de Fisica Aplicada II, \\ Universidad 
del Pais Vasco, Apartado 644,
             48080 Bilbao, Spain \\
        $^{3}$  Center for Materials Theory,
                   Department of Physics and Astronomy,\\
         Rutgers University, Piscataway, New Jersey 08855-0849, USA}

\date{February 14, 2001}

\maketitle

\begin{abstract}

A first-principles-derived approach is used to study structural,
piezoelectric and dielectric properties of
Pb(Zr$_{1-x}$Ti$_{x}$)O$_{3}$ (PZT) solid solutions near the
morphotropic phase boundary at low temperature.  Three ferroelectric
phases are found to exist: a tetragonal phase for larger $x$
compositions, a rhombohedral phase for smaller $x$ compositions, and
the recently discovered monoclinic phase in between. In this
monoclinic phase, the polarization associated with the Zr atoms
behaves differently from the polarization associated with the Ti
atoms.  As the composition $x$ decreases, the former rotates more
quickly towards the pseudo-cubic $\lbrack 111 \rbrack$ direction and
grows in magnitude, while the latter lags in its rotation and its
magnitude shrinks.  The local microscopic structure is found to
deviate significantly from the average structure in these PZT alloy
phases as a result of fluctuations in the directions and magnitudes of
the local polarizations.  The monoclinic phase is characterized by a
very large piezoelectric and dielectric response.

\end{abstract}
\vspace{0.3cm}



\vskip1pc
]
\narrowtext \marginparwidth 2.7in \marginparsep 0.5in

\section{Introduction}
Since the beginning of the 70's, Pb(Zr$_{1-x}$Ti$_{x}$)O$_{3}$ (PZT)
alloys have been known to exhibit a morphotropic phase boundary (MPB)
separating a ferroelectric region with a tetragonal ground state ($x$
$ >$ 0.52) from a ferroelectric region with rhombohedral symmetry ($x$
$<$ 0.45) \cite{Jaffe}.  Very recently, synchrotron x-ray powder
diffraction studies have revealed that there is in fact a third phase
in the vicinity of the MPB at low temperature \cite{Noheda1}. This
phase is ferroelectric, adopts a monoclinic symmetry and occurs within
a narrow range of composition $x$ that separates the tetragonal region
from the rhombohedral region \cite{Noheda1,Noheda2}.  It has been
suggested that this monoclinic phase acts as a transitional bridge
between the tetragonal phase, for which the electrical polarization
$\bf{P}$ lies along the pseudo-cubic [001] direction, and the
rhombohedral phase, for which $\bf{P}$ is along the pseudo-cubic [111]
direction \cite{Noheda1}.  The use of a recently developed
first-principles approach has confirmed that this is indeed the case,
since the polarization of the monoclinic phase has been found to
rotate from the [001] to the [111] direction as the composition $x$
decreases in the MPB region \cite{Paper1}.

Despite these experimental and theoretical advances, some features
related to this MPB are still unknown.  For instance, one may wonder
what is the separate contribution of Zr and Ti atoms to the rotation
of the polarization in this monoclinic phase.  One may also want to
know if the local (microscopic) structure of a PZT alloy near its MPB
differs strongly from its average (macroscopic) structure, as
suggested by Ref. \onlinecite{Noheda2}.

Moreover, PZT is strongly piezoelectric in the vicinity of the MPB
\cite{Berlincourt}, which is a primary reason why these materials are
of current use in transducers and other piezoelectric devices
\cite{Uchino}.  Recent theoretical and experimental work has
reexamined the piezoelectricity of tetragonal and rhombohedral PZT
alloys \cite{Paper1,Noheda3}.  However, we are not aware of any
previous studies of the piezoelectric or dielectric response in the
monoclinic phase of PZT.

The purpose of the present article is to apply the newly-developed
first-principles derived scheme of Ref. \onlinecite{Paper1} to PZT and
to investigate all the issues mentioned above.  Our main findings are
as follows. (1) As the Ti composition decreases in the recently
discovered monoclinic phase, the polarization associated with Zr atoms
rotates towards the [111] direction and grows in magnitude, while the
polarization associated with Ti atoms lags in its rotation toward this
[111] direction and shrinks in magnitude. (2) There are strong
fluctuations of the direction and magnitude of the local polarizations
centered in the different unit cells of PZT alloys near the MPB,
indicating that various unit cells adopt a local structure somewhat
different from the macroscopic average. (3) The recently discovered
monoclinic phase exhibits very large piezoelectric and dielectric
coefficients.

The paper is organized as follows. In Sec. II, we describe our
approach in detail. Section III reports our results, and we conclude
in Sec. IV.

\section{Methods}

We use the numerical scheme proposed in Ref. \onlinecite{Paper1},
which consists of constructing an alloy effective Hamiltonian from
first-principles calculations.  For a {\it ferroelectric} material,
the effective Hamiltonian should include structural degrees of freedom
corresponding to the ferroelectric local soft mode and the strain
variables.  These are the most important degrees of freedom because
ferroelectric transitions are accompanied by a softening of the phonon
soft mode and by the appearance of a strain \cite{ZhongDavid}.
Moreover, a realistic {\it alloy} effective Hamiltonian must also
include the compositional degrees of freedom, because the atomic
arrangement can strongly affect the ferroelectric properties of an
alloy \cite{Setter}.  We proposed to incorporate all such degrees of
freedom by writing the total energy $E$ as a sum of two energies,
\begin{eqnarray}
\lefteqn{\nonumber
   E (\{ { \bf u_{\it i}} \}, \{ { \bf v_{\it i}} \}, \eta_{\it H} 
      \{ \sigma_{\it j} \}) =}\\ \nonumber
   & & E_{\rm ave} 
       (\{ { \bf u_{\it i}} \}, \{ { \bf v_{\it i}} \}, \eta_{\it H}) \\
   & & \mbox{} +
   E_{\rm loc} (\{ { \bf u_{\it i}} \},\{ { \bf v_{\it i}} \},
   \{ \sigma_{\it j} \}) \;\;,
\end{eqnarray}
where ${\bf u_{\it i}}$ is the local B-centered soft mode in unit cell
$i$ of the ABO$_{3}$ perovskite material under study; $\{ { \bf v_{\it
i}} \}$ are the dimensionless local displacements which are related to
the inhomogeneous strain variables inside each cell and which are
centered on the A-sites \cite{ZhongDavid}; $\eta_{\it H}$ is the
homogeneous strain tensor; and the $\{ \sigma_{{\it j}}\}$
characterize the atomic configuration of the alloy. That is,
$\sigma_{\it j}$=+1 or $-1$ corresponds to the presence of a Zr or Ti
atom, respectively, at the B-sublattice site $j$ of the
Pb(Zr$_{1-x}$Ti$_{x}$)O$_{3}$ solid solution.  The energy $E_{\rm
ave}$ depends only on the soft-mode and strain variables.  The $\{
\sigma_{{\it j}}\}$ parameters are incorporated into the second energy
term $E_{\rm loc}$, which thus accounts for the chemical differences
between Zr and Ti atoms.

Expressions for the total energy $E$ have recently been proposed
\cite{ZhongDavid,Rabe} for {\it simple} ABO$_{3}$ perovskite systems
(i.e., in the absence of $\{\sigma_{{\it j}}\}$ variables). These have
been very successful both for reproducing phase transition sequences
\cite{ZhongDavid,Rabe,Krakaeur} and for studying ferroelectric domain
walls \cite{Jorge}, as well as for calculating finite-temperature
dielectric and electromechanical properties
\cite{Alberto1,Alberto2,Cockayne}.  Here, for $E_{\rm ave}$, we
generalized the analytical expression of Ref.~\onlinecite{ZhongDavid}
to the case of the Pb(Zr$_{1-x}$Ti$_{x}$)O$_{3}$ alloy, by making use
of the virtual crystal approximation (VCA)
\cite{VCA,LaurentDavid3,Rappe}.  We thus replaced the
Pb(Zr$_{1-x}$Ti$_{x}$)O$_{3}$ alloy by a virtual (uniform)
Pb$\langle$B$\rangle$O$_{3}$ simple system in which the
$\langle$B$\rangle$ atom is a virtual atom involving a kind of
potential average between Zr and Ti atoms \cite{LaurentDavid3}.
$E_{\rm ave}$ thus consists of five parts: a local-mode self-energy, a
long-range dipole-dipole interaction, a short-range interaction
between soft modes, an elastic energy, and an interaction between the
local modes and local strain \cite{ZhongDavid}.  The analytical
expression for $E_{\rm ave}$ has 18 free parameters that are
determined by fitting to the results of almost 40 first-principles
calculations on small VCA cells (typically between 5 and 20
atoms/cell) following the procedure of Ref.~\onlinecite{ZhongDavid}.
More precisely, the first-principles method used is the plane-wave
ultrasoft-pseudopotential method \cite{USPP} within the local-density
approximation (LDA) \cite{LDA}, and the VCA approach adopted is the
one of Ref.~\onlinecite{LaurentDavid3}. Table I reports the resulting
18 expansion parameters of $E_{\rm ave}$ for the
Pb(Zr$_{0.5}$Ti$_{0.5}$)O$_{3}$ alloy.

\begin{table}[h!]
\caption{Expansion parameters (in atomic units) of $E_{\rm ave}$ for
 Pb(Zr$_{0.5}$Ti$_{0.5}$)O$_{3}$.}
\label{Table I}
\begin{tabular}{ld}
  Parameter & Value \\
\tableline
~~$\kappa_{2}$   &   0.0138 \\
~~$\alpha$ & 0.011\\
~~$\gamma$ & 0.002\\
\tableline
~~$j_{1}$ &  $-$0.00577 \\
~~$j_{2}$ &   0.01425 \\
~~$j_{3}$ &   0.00140 \\
~~$j_{4}$ &  $-$0.00094\\
~~$j_{5}$ &   0.00141\\
~~$j_{6}$ &   0.00006\\
~~$j_{7}$ &   0.00003\\
\tableline
~~$B_{11}$ &  5.22\\
~~$B_{12}$ &  1.67\\
~~$B_{44}$ &  1.22\\
\tableline
~~$B_{1xx}$ & $-$0.374\\
~~$B_{1yy}$ & $-$0.155\\
~~$B_{4yz}$ & $-$0.068\\
\tableline
~~$Z^{*}$ & 7.342\\
~~$\epsilon_{\infty}$ & 7.150\\
\end{tabular}
\end{table}

While expressions were available for $E_{\rm ave}$, we were not aware
of any analytical expression that had previously been proposed and
tested for $E_{\rm loc}$.  Following the spirit of the ``computational
alchemy'' method developed for calculating the compositional energy of
semiconductor alloys \cite{Gironcoli,Nicola,Marco}, we derived E$_{\rm
loc}$ by treating the alloy configuration $\{ \sigma_{j}\}$ as a
perturbation of the VCA system.  We adopted an expression that
includes: (i) the {\it on-site} effect of alloying on the self-energy
up to the fourth order in the local mode amplitude $\bf{u_{\it i}}$;
and (ii) the {\it intersite} contributions involving the first-order
terms in a perturbation expansion in powers of $\sigma_{\it j}$ (i.e.,
terms that are linear in $\bf{u_{\it i}}$ or $\bf{v_{\it i}}$).  That
is,
\begin{eqnarray}
  E_{\rm loc} && (\{ { \bf u_{\it i}} \},\{ { \bf v_{\it i}} \},
  \{ \sigma_{\it j} \}) = \nonumber \\
  && \sum_{i} [ \Delta \alpha (\sigma_{\it i}) ~ u_{\it i}^{4} ~+
  ~ \Delta \gamma (\sigma_{\it i}) ~( u_{\it ix}^{2}u_{\it iy}^{2} +
  u_{\it iy}^{2}u_{\it iz}^{2} + u_{\it iz}^{2} u_{\it ix}^{2})]\nonumber \\
  && +~ \sum_{ij}
  [Q_{\it |j-i|}~\sigma_{\it j}~ { \bf e_{\it ji}} \cdot { \bf u_{\it i}}~+~
  R_{\it |j-i|}~\sigma_{\it j}~ { \bf f_{\it ji}} \cdot { \bf v_{\it i} }]
  \;\;,
\end{eqnarray}
where the sum over $i$ runs over all the unit cells, while the sum
over $j$ runs over the mixed sublattice sites.  $u_{\it ix}$, $u_{\it
iy}$ and $u_{\it iz}$ are the Cartesian coordinates of the local-mode
$\bf{u_{\it i}}$.  $\bf{e_{\it ji}}$ is a unit vector joining the site
$j$ to the center of the soft mode $\bf{u_{\it i}}$, and $\bf{f_{\it
ji}}$ is a unit vector joining the site $j$ to the origin of
$\bf{v_{\it i}}$.  $\Delta \alpha (\sigma_{\it i})$ and $\Delta \gamma
(\sigma_{\it i})$ characterize the on-site contribution of
alloying. Their strength and sign reflect how the identity of the atom
sitting on the $i$ site affects the local-mode self energy of the
``VCA'' Pb(Zr$_{1-x}$Ti$_{x}$)O$_{3}$ solid solution.  $Q_{{\it
|j-i|}}$ and $R_{{\it |j-i|}}$ are related to intersite interactions
between the alloy parameter $\sigma_{\it j}$ on the site $j$ and the
local mode $u_{\it i}$ and the strain-related $v_{\it i}$ at the site
$i$, respectively.  $Q_{{\it |j-i|}}$ and $R_{{\it |j-i|}}$ only
depend on the distance between $i$ and $j$ up to the third neighbors
shell, while for symmetry reasons, the expression for the intersite
interactions becomes more complex when going beyond the third
neighbors shell.  In principle, terms involving higher powers of $\{
\sigma_{\it j} \}$, $\bf{u_{\it i}}$ and $\bf{v_{\it i}}$ might be
included to improve the quality of the expansion, but as shown in Ref.
\onlinecite{Paper1}, we found this level of expansion to give a very
good account of experimental findings.  We also found that $Q_{\it
|j-i|}$ and $R_{\it |j-i|}$ rapidly decrease as the distance between
$i$ and $j$ increases. As a result, we included the contribution up to
the third neighbors for $Q_{\it |j-i|}$ (denoted $Q_{\it 1}$, $Q_{\it
2}$ and $Q_{\it 3}$ in the following) and up to the first neighbor
shell for $R_{\it |j-i|}$ (denoted $R_{\it 1}$).

The parameters $\Delta \alpha (\sigma_{\it i})$, $\Delta \gamma
(\sigma_{\it i})$, $Q_{{\it |j-i|}}$ and $R_{{\it |j-i|}}$ are also
derived by performing first-principles calculations, and are given in
Table II in the case of the Pb(Zr$_{0.5}$Ti$_{0.5}$)O$_{3}$ alloy.
More precisely, $\Delta \alpha (\sigma_{\it i})$ and $\Delta \gamma
(\sigma_{\it i})$ are derived by computing the energy of a 5-atom cell
containing a true B-atom [e.g., Ti or Zr in Pb(Zr,Ti)O$_{3}$] when the
atoms are displaced as in the VCA local mode.  Then $Q_{{\it |j-i|}}$
and $R_{{\it |j-i|}}$ are derived by using large ideal cubic cells (up
to 40 atoms) containing a central true B-atom surrounded by VCA atoms,
and are simply related to the atomic forces occurring on these VCA
atoms.

\begin{table}[h!]
\caption{Expansion parameters of $E_{\rm loc}$ (in atomic units)
for  Pb(Zr$_{0.5}$Ti$_{0.5}$)O$_{3}$.
 $\sigma$=+1 and $-1$ corresponds to the
presence of a Zr and Ti atom, respectively.}
\label{Table II}
\begin{tabular}{ld}
  Parameter & Value \\
\tableline
~~$\Delta \alpha (+1)$ & 0.003\\
~~$\Delta \alpha (-1)$ & $-$0.003\\
~~$\Delta \gamma (+1)$ & $-$0.010\\
~~$\Delta \gamma (-1)$ & 0.003\\
\tableline
~~$Q_{1}$ &   0.00160 \\
~~$Q_{2}$ &  $-$0.00028 \\
~~$Q_{3}$ &  $-$0.00018 \\
\tableline
~~$R_{1}$ &  $-$0.0125\\
\end{tabular}
\end{table}

In principle, all the parameters in Eqs. (1) and (2) should change
when one varies the composition $x$ in the
Pb(Zr$_{1-x}$Ti$_{x}$)O$_{3}$ solid solution.  However, we assumed
that only the parameters related to the local-mode self-energy --
i.e., $\alpha$ and $\gamma$ for the VCA alloy, and $\Delta \alpha
(\sigma_{\it i})$ and $\Delta \gamma (\sigma_{\it i})$ in Eq (2) --
can significantly change with composition.  This
composition-dependence was assumed to be linear, and was determined by
performing first-principles simulations on cells with two different
compositions, namely $x$=0.5 and $x=0.45$.  The resulting
composition-dependencies are
\begin{eqnarray}
&& \alpha~+~\Delta \alpha (+1)~=~\phantom{-}0.014~+~ 0.02~(0.5-x) 
  \nonumber \\
 && \alpha~+~\Delta \alpha (-1)~=~\phantom{-}0.008~+~0.02~(0.5-x) 
  \nonumber \\
&&  \gamma~+~\Delta \gamma (+1)~=~          -0.008~-~0.14~(0.5-x) 
  \nonumber \\
&&   \gamma~+~\Delta \gamma (-1)~=~\phantom{-}0.005~-~0.02~(0.5-x) 
  \nonumber \\
\end{eqnarray}
Such a linear composition-dependence approach is only realistic
when exploring a narrow range of compositions, as done in Ref.
\onlinecite{Paper1} and in the present study.

Once our effective Hamiltonian is fully specified, the total energy of
Eq.(1) is used in Monte-Carlo simulations to compute
finite-temperature properties of PZT alloys.  We use $10\times
10\times 10$ supercells (5000 atoms), since this choice yields
well-converged results at low temperature \cite{Paper2}.  The $\{
\sigma_{\it j} \}$ variables are chosen randomly in order to mimic
maximal compositional disorder, consistent with experimental reality
\cite{Cross}, and are kept fixed during the Monte-Carlo simulations.
We find that averaging our results over a couple of different
realizations of the disorder leads to well-converged statistical
properties.  The outputs of the Monte-Carlo procedure are the local
mode vectors ${\bf u}$ and the homogeneous strain tensor $ \eta_{\it
H}$.  We use the correlation-function approach of
Refs~\onlinecite{Alberto1,Alberto2} to derive the piezoelectric and
dielectric response from these Monte-carlo simulations.  Up to
10$^{6}$ Monte-Carlo sweeps are first performed to equilibrate the
system, and then 2$\times$10$^{4}$ sweeps are used to get the various
statistical averages.  In the present study, the temperature is kept
fixed at 50\,K. Note that Refs. \onlinecite{Paper1,Paper2} demonstrate
that our approach leads to a (converged) theoretical Curie temperature
$T_{c,theo}$ of 1032\,K for Pb(Zr$_{0.5}$Ti$_{0.5}$)O$_{3}$, which is
much higher than the experimental value $T_{c,exp}$ of 640\,K
\cite{YamamotoJJ}. This difficulty of reproducing $T_{c}$ seems to be
a general feature of the effective-Hamiltonian approach
\cite{ZhongDavid,Rabe,Krakaeur}, and may be due to higher perturbative
terms not included in the analytical expression for the total energy.
When comparing with measurements, this shortcoming can be overcome by
multiplying the temperature used in the simulation by a constant
factor of $T_{c,exp}$/$T_{c,theo}$ \cite{Paper1,Alberto1}.  As a
result, our simulated temperature of 50\,K corresponds to an
``experimental'' temperature around 30\,K.

\section{Results}

\subsection{Structural properties}

The averaged homogeneous strain variables obtained in
Pb(Zr$_{1-x}$Ti$_{x}$)O$_{3}$ from our simulations are shown in Fig. 1
as a function of the composition $x$. These strain variables are
measured relative to the LDA-calculated minimum-energy cubic structure
with lattice constant $a_{0}=7.56$ a.u., and are expressed in the
Voigt notation.  For Ti compositions larger than 49\%, we have
$\eta_1=\eta_2 \neq 0$, $\eta_3$ $>$ $\eta_2$, and
$\eta_4=\eta_5=\eta_6=0$. This strain tensor corresponds to a
tetragonal phase with 5 atoms per unit cell.
\begin{figure}
\epsfxsize=\hsize\epsfbox{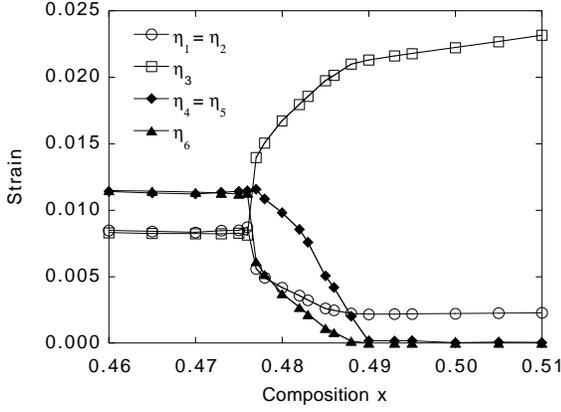}
\caption{The average homogeneous strain tensor $\eta_{H}$ as a
function of composition in disordered Pb(Zr$_{1-x}$Ti$_{x}$)O$_{3}$
at 50\,K.  Strains are measured relative to the theoretical
minimum-energy cubic structure of lattice constant 7.56 a.u.}
\end{figure}
For compositions lower than 47.5\%, we predict that the system adopts
the so-called ``high-temperature'' (5-atoms per unit cell)
rhombohedral phase \cite{Lines} since $\eta_1=\eta_2=\eta_3 \neq 0$
and $\eta_4=\eta_5=\eta_6 \neq 0$.  The most interesting feature of
Fig.~1 is the behavior of the homogeneous strain for the compositional
range between 47.5\% and 49\%: as $x$ decreases, (1) $\eta_{1}$ and
$\eta_{2}$ slightly increase and remain equal to each other, (2)
$\eta_3$ strongly decreases but is still larger than $\eta_{1}$, (3)
$\eta_{4}$ and $\eta_{5}$ increase and are equal to each other, and
(4) $\eta_{6}$ increases and is smaller than $\eta_{4}$.  This
behavior is characteristic of an intermediate phase that is neither
tetragonal nor rhombohedral, but rather is monoclinic.  From the
strain tensor shown in Fig.~1, we further predict that this monoclinic
phase can be characterized by a 10-atom conventional cell (5-atom
primitive cell) with lattice vectors ${\bf a}_m = a_0
(-1-\eta_1-\eta_6/2,\,-1-\eta_1-\eta_6/2,-\eta_4)$, ${\bf b}_m = a_0
(1+\eta_1-\eta_6/2,\,-1-\eta_1+\eta_6/2,0)$, and ${\bf c}_m = a_0
(\eta_4/2,\,\eta_4/2,1+\eta_3)$.
These predictions are in excellent quantitative agreement with the
lattice vectors of the monoclinic phase experimentally discovered by
Noheda {\it et al.} \cite{Noheda1}.  Fig~1 also reveals that the
strain variables continuously change with composition when crossing
the tetragonal--to--monoclinic transition, while these variables
exhibit a sudden jump at the monoclinic--to--rhombohedral transition
composition.  Interestingly, these two distinct features are
consistent with the predictions of Ref. \onlinecite{DavidMorrel},
i.e., that the tetragonal--to--monoclinic transition is of second
order while the monoclinic--to--rhombohedral transition is of
first-order.

Our simulations also agree with measurements for the narrowing of the
compositional range of the monoclinic phase when increasing the
temperature \cite{Noheda2}. For instance, Fig.~1 indicates that the
monoclinic phase is predicted to occur for $0.475 < x < 0.49$ at
$T$=50\,K, while we predict (not shown here) that increasing the
simulated temperature up to 485K -- which corresponds to an
experimental temperature of 300\,K -- leads to the existence of the
monoclinic phase for $0.475 < x < 0.485$. As observed in
Ref.\onlinecite{Noheda2}, the tetragonal--to--monoclinic transition
composition thus decreases when increasing temperature while the
rhombohedral--to--monoclinic transition concentration is independent
of the temperature.  Unfortunately, the calculations are not
sufficiently precise to determine whether the phase diagram of PZT
exhibits a triple point where the rhombohedral, monoclinic and
tetragonal phases meet at a given composition and temperature as
suggested in Ref. \onlinecite{Noheda4}, or whether the monoclinic
phase instead survives right up to the boundary with the paraelectric
cubic phase. Furthermore, Ref. \onlinecite{Noheda4} reports that, at
$T$=20\,K, the monoclinic phase is observed for $0.46 < x < 0.51$ ,
i.e. for a larger range that the one shown in Fig. 1.  This may be
explained by our empirical finding that the range of the monoclinic
phase depends strongly on the parameters of Eq.~(3). A slight
adjustment of these parameters may thus lead to a better agreement
with experiment for the compositional range of the monoclinic phase.
Note also that our model cannot predict the so-called
``low-temperature'' rhombohedral phase \cite{Lines} since this
involves oxygen octahedra-tilting degrees of freedom that are not
including in the present model.  Consequently, our simulations
demonstrate that the oxygen tilts are not the driving force for the
transition or for the occurrence of the monoclinic phase.

\begin{figure}
\epsfxsize=\hsize\epsfbox{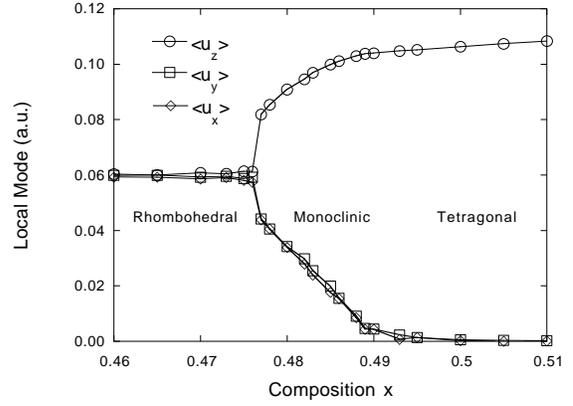}
\caption{Average cartesian coordinates $\langle u_{x}\rangle$,
$\langle u_{y}\rangle$ and $\langle u_{z}\rangle$ of the
local-mode vector as a function of composition $x$ in disordered
Pb(Zr$_{1-x}$Ti$_{x}$)O$_{3}$ at 50\,K, corresponding to
an experimental $T$=30\,K after rescaling (see text).}
\end{figure}
Figure 2 shows the cartesian coordinates ($\langle u_{x}\rangle$,
$\langle u_{y}\rangle$ and $\langle u_{z}\rangle$) of the supercell
average of the local mode vectors in Pb(Zr$_{1-x}$Ti$_{x}$)O$_{3}$ as
a function of the composition $x$ at $T=50\,K$, as predicted by our
approach described by Eqs.~(1) and (2).  The average local mode
$\langle {\bf u}\rangle$, and hence the polarization, is parallel to
the pseudo-cubic [001] direction in the tetragonal phase ($x >0.49$\%)
while the polarization becomes parallel to the pseudo-cubic [111]
direction in the rhombohedral phase ($x< 0.475$\%).  In the monoclinic
phase, the change of strain shown in Fig.~1 is associated with a
decrease of $\langle u_{z}\rangle$ while $\langle u_{x}\rangle$ and
$\langle u_{y}\rangle$ increase and remain nearly equal to each other
as the composition $x$ decreases from 49\% to 47.5\%.  Figure~2 thus
demonstrates that the electrical polarization rotates from the
pseudo-cubic [001] direction to the the pseudo-cubic [111] direction
as the Ti composition decreases in this monoclinic phase.

\begin{figure}
\epsfxsize=\hsize\epsfbox{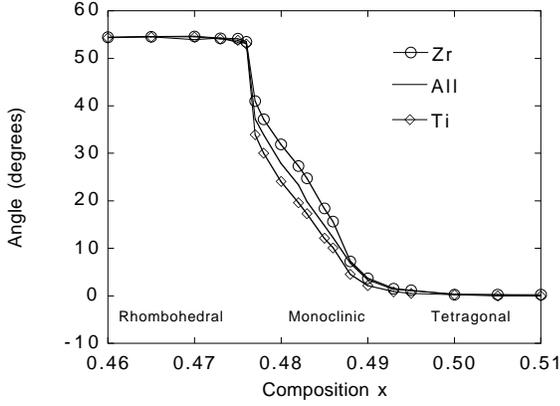}
\caption{Angle between the average local-mode vector and the
pseudo-cubic [001] direction as a function of composition in
disordered Pb(Zr$_{1-x}$Ti$_{x}$)O$_{3}$ at 50\,K.
Circles and diamonds refer to averages over local modes centered
on Zr and Ti sites, respectively; solid line refers to average over
all local modes.}
\end{figure}
\begin{figure}
\epsfxsize=\hsize\epsfbox{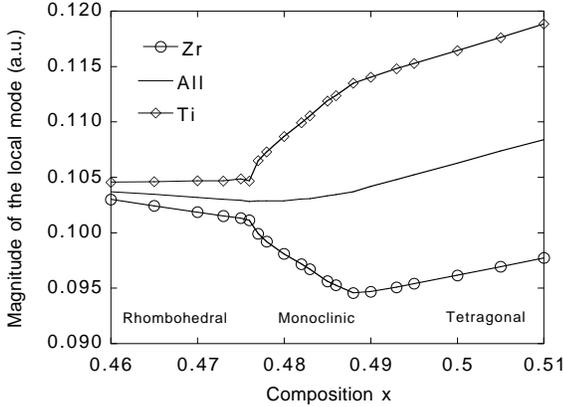}
\caption{Magnitude of the average local-mode vector as a function
of composition in
disordered Pb(Zr$_{1-x}$Ti$_{x}$)O$_{3}$ at 50\,K.
Circles and diamonds refer to averages over local modes centered
on Zr and Ti sites, respectively; solid line refers to average over
all local modes.}
\end{figure}

To better understand the separate contribution of Zr and Ti atoms to
this rotation, Fig.~3 shows the composition dependence of the angle
$\delta$ between the pseudo-cubic [001] direction and the ${\bf
u}_{\rm Zr}$ (respectively, ${\bf u}_{\rm Ti}$) local mode vectors
averaged over cells centered on Zr (respectively, Ti) atoms. The angle
between the pseudo-cubic [001] direction and the entire supercell
average $\langle{\bf u}\rangle$ is also shown in Fig.~3.  Similarly,
Fig.~4 displays the variation of the magnitude of $\langle{\bf
u}\rangle$, ${\bf u}_{\rm Zr}$ and ${\bf u}_{\rm Ti}$ as a function of
the composition.  Figure 3 demonstrates that, in the monoclinic phase,
${\bf u}_{\rm Zr}$ and ${\bf u}_{\rm Ti}$ also rotate from the [001]
direction -- for which $\delta=0$ -- to the [111] direction -- for
which $\delta=54.7^\circ$ -- as the Ti composition decreases from 49\%
to 47.5\%.  In this monoclinic phase, the pseudo-cubic [111] direction
is always closer to ${\bf u}_{\rm Zr}$ than to ${\bf u}_{\rm Ti}$.
Furthermore, Fig.~4 indicates that the magnitude of ${\bf u}_{\rm Zr}$
is smaller than the magnitude of ${\bf u}_{\rm Ti}$ in the entire
monoclinic phase concentration range, and that the magnitude of ${\bf
u}_{\rm Zr}$ increases while the magnitude of ${\bf u}_{\rm Ti}$
decreases when the Ti composition decreases from 49\% to 47.5\%. This
leads to a nearly composition-independent magnitude of $\langle{\bf
u}\rangle$.  In other words, the total polarization simply rotates in
the monoclinic phase while the polarization associated with Zr
(respectively, Ti) atoms rotates and also grows (respectively,
shrinks), as the Ti concentration decreases.

The authors of Ref. \onlinecite{Noheda2} suggested that near the MPB,
the rhombohedral and tetragonal phases of PZT can be described in
terms of a structure that is locally monoclinic, and in which the
average polarization along the [111] or [001] pseudo-cubic direction
can occur by means of fluctuations between a subset of three or four
nearby monoclinic orientations.  In this picture, the transition from
the rhombohedral or tetragonal phase to the monoclinic phase would
occur by the freezing in of one of these monoclinic orientations.  Our
effective Hamiltonian can be used to investigate the local structure
of PZT around its MPB, and, in particular, to check if the local
structure of the tetragonal or rhombohedral phase is different from
its average structure.  Figures 5(a-c) display the predicted local
modes distributions, at T=50\,K, for Pb(Zr$_{1-x}$Ti$_{x}$)O$_{3}$
solid solutions with $x$=0.50 (tetragonal average structure),
$x$=0.482 (monoclinic average structure) and $x$=0.47 (rhombohedral
average structure), respectively.  One can clearly see that the
direction and magnitude of the local modes fluctuate around their
average value in any of these three PZT solid solutions. As a result,
various unit cells adopt a local structure different from the
macroscopic average.  However, our calculations do not support the
hypothesis that the tetragonal and rhombohedral phases of PZT are
simply made up of a small number of local monoclinic phases, since
Figs.~5(a) and 5(c) do not show the distributions breaking up into
clusters centered along the monoclinic directions.
 
\begin{figure}
\begin{center}
\epsfxsize=6truecm\epsfbox{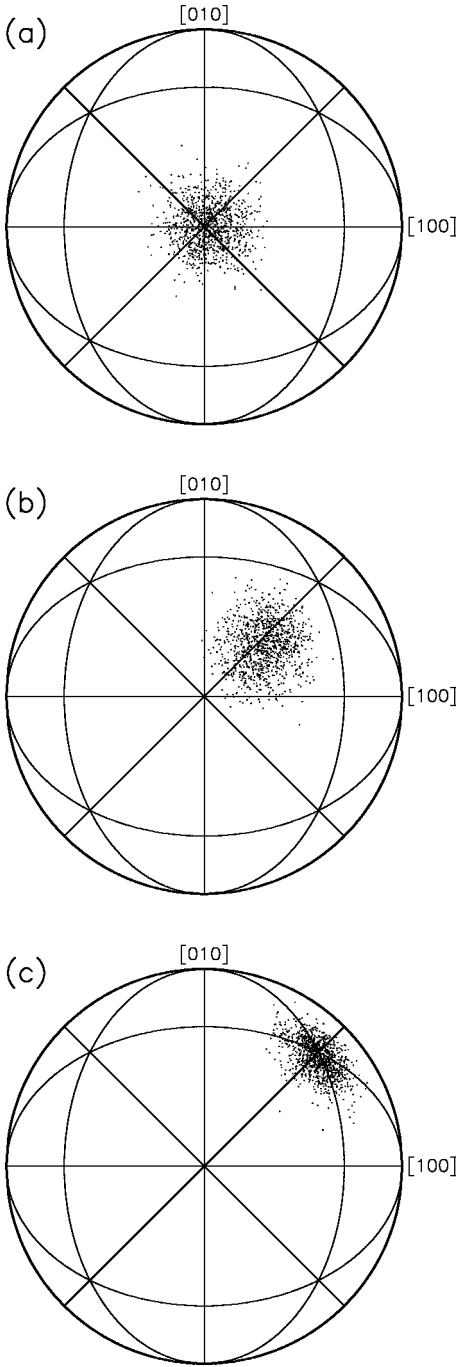}
\end{center}
\vspace{0.3truecm}
\caption{Projection of the distribution of local-mode orientations for
disordered Pb(Zr$_{1-x}$Ti$_{x}$)O$_{3}$ at 50\,K for (a) $x$=0.50,
(b) $x$=0.482, and (c) $x$=0.47.}
\end{figure}

\subsection{Piezoelectric and dielectric properties}

We now use our alloy effective Hamiltonian to investigate the
piezoelectric and dielectric properties of
Pb(Zr$_{1-x}$Ti$_{x}$)O$_{3}$ {\it vs.} $x$ at T = 50\,K.  Figure 6
shows the piezoelectric coefficients $d_{33}$ and $d_{15}$ as a
function of the composition $x$, when representing the piezoelectric
tensor in the orthonormal basis formed by {\bf a$_{1}$} = [100], {\bf
a$_{2}$} = [010] and {\bf a$_{3}$} = [001].  Figure 7 displays the
composition dependence of the dielectric susceptibilities $\chi_{11}$
and $\chi_{33}$ expressed in the same basis.

One can notice that $d_{15}$ is much larger than $d_{33}$ in the
tetragonal phase, i.e., for $x>49$\%.  This is consistent with recent
measurements revealing that the piezoelectric elongation of the
tetragonal unit cell of PZT does not occur along the polar [001]
direction \cite{Noheda3}.  The large value of $d_{15}$ also explains
the strong piezoelectric response observed in ceramic samples, since
this latter involves an average over the single-crystal coefficients
$d_{15}$ and $d_{33}$ \cite{Paper1}.

\begin{figure}
\epsfxsize=\hsize\epsfbox{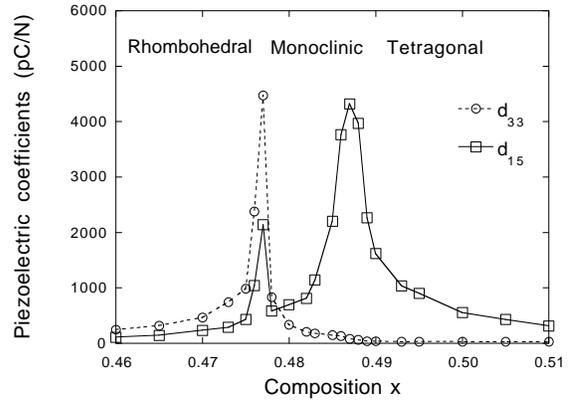}
\caption{Piezoelectric coefficients d$_{33}$ and d$_{15}$ as a
function of composition in disordered Pb(Zr$_{1-x}$Ti$_{x}$)O$_{3}$
at 50\,K. Statistical errors
are estimated to be $\sim$10\% of the values displayed.}
\end{figure}

One can also see that $d_{33}$ is predicted be quite large in the
rhombohedral phase (occurring for $x<47.5$\%). This prediction agrees
with the theoretical and experimental findings of
Ref. \onlinecite{Du,Park} that the $d_{33}$ coefficient of
rhombohedral materials can be very large along the [001] direction,
i.e. away from the polar direction which is oriented along the
pseudo-cubic [111] direction.

 Fig.~6 also reveals that $d_{33}$ reaches its largest value near the
monoclinic--to--rhombohedral transition, while $d_{15}$ peaks near
both the monoclinic--to--rhombohedral and tetragonal--to--monoclinic
transitions.  The most striking feature of Fig.~6 is that $d_{15}$ has
a remarkably large value -- above 600 pC/N -- in the entire monoclinic
phase range.

Fig.~7 demonstrates that the dielectric susceptibility $\chi_{33}$
behaves in a similar way as the piezoelectric coefficient $d_{33}$, in
the sense that $\chi_{33}$ is also peaked near the
monoclinic--to--rhombohedral transition and is also much larger in the
rhombohedral phase than in the tetragonal structure.  On the other
hand, $\chi_{11}$ behaves in a manner similar to $d_{15}$, since they
both have peaks at both transitions, and are both very large in the
monoclinic phase.

\begin{figure}
\epsfxsize=\hsize\epsfbox{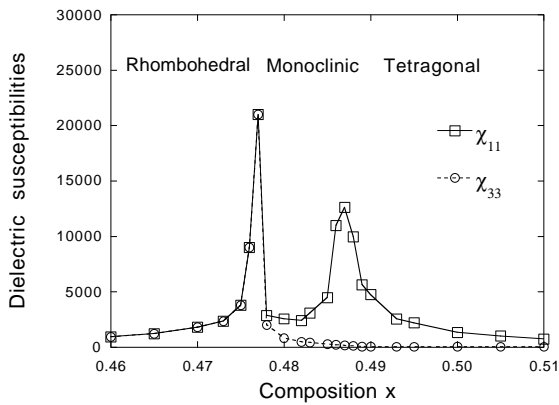}
\caption{Dielectric susceptibilities $\chi_{11}$ and $\chi_{33}$
as a function of composition in disordered Pb(Zr$_{1-x}$Ti$_{x}$)O$_{3}$
at 50\,K. Statistical errors
are estimated to be $\sim$10\% of the values displayed.}
\end{figure}

It thus appears that the rotation of the polarization (see Fig.~2) not
only leads to a very high $d_{15}$ piezoelectric response (see
Fig.~6), consistent with the finding of Ref. \onlinecite{Fu}, but also
to a large $\chi_{11}$ dielectric response (see Fig.~7).

\section{Conclusions}

In summary, we have used the first-principles derived computational
scheme proposed in Ref. \onlinecite{Paper1} to study low-temperature
properties of disordered Pb(Zr$_{1-x}$Ti$_{x}$)O$_{3}$ solid solutions
near the MPB.

We find that the monoclinic phase reported in
Ref. \onlinecite{Noheda1} acts as a structural bridge between the
tetragonal phase ($x>49$\% at $T$=50\,K) and the rhombohedral phase
($x<47.5$\% at $T$=50\,K), with the electric polarization rotating
from the pseudo-cubic [001] to [111] direction as the Ti concentration
decreases through the monoclinic range.  The polarization associated
with the Zr atoms differs from the polarization associated with Ti
atoms in the monoclinic phase, in that the former rotates faster
towards the [111] direction and grows in magnitude, while the latter
rotates more slowly and shrinks as the composition $x$ decreases.
Furthermore, we also investigated the local structures of tetragonal,
monoclinic and rhombohedral PZT alloys, and found that the local
polarizations centered in the different unit cells fluctuate
significantly in both magnitude and direction around their average
value.  However, the pattern of these variations does not support a
picture in which the tetragonal or rhombohedral phases could be
regarded as arising from fluctuations among neighboring monoclinic
states.  Finally, some piezoelectric and dielectric coefficients are
predicted to be extremely large in this monoclinic phase.

\section{Acknowledgments}

L.B.~thanks the financial assistance provided by the Arkansas Science
and Technology Authority (grant N99-B-21), the Office of Naval
Research (grant N00014-00-1-0542) and the National Science Foundation
(grant DMR-9983678).  A.G.\ acknowledges support from the Spanish
Ministry of Education (grant PB98-0244).  D.V.\ acknowledges the
financial support of Office of Naval Research grant N00014-97-1-0048.
We wish to thank B. Noheda and T. Egami for very useful discussions.

\end{document}